# Battery Material Comparisons Should Refocus on Diffusivity with Best Practices

**CJ Sturgill[1], Roya Rajabi[1], Md Abdullah Al Muhit[1], Hans-Conrad zur Loye[1], Morgan Stefik[1*]**

1) Department of Chemistry and Biochemistry, University of South Carolina, Columbia, SC 29208.

*Corresponding Author: morgan@stefikgroup.com

**Abstract**

The continuous demand for improved batteries motivates the discovery and advancement of materials with improved transport. Ionic diffusivity is the relevant material property where its measurement depends on accurate assessment of the active material length-scale, generally from the mass-specific surface area. In this perspective, we argue for renewed focus on diffusivity comparisons. A procedural review of 303 recent open-access publications about battery material development revealed two aspects: (1) ~49% of publications support structure-property transport claims using diffusivity values and (2) of those reporting diffusivity values, 15% clearly stated that the length scale was measured after grinding-alone or stated that grinding was not used at all. Diffusivity assessment rationally requires length scale measurement after grinding ("grind-measure"), rather than the reverse. A range of measurement methods are compared, including SEM, BET, and SAXS as well as the resulting apparent diffusivities. Common errors and pitfalls of each of these approaches are described. As an example, datasets are presented for $TiNb_2O_7$ (TNO1) and $Ti_2Nb_{10}O_{29}$ (TNO2) made from sol-gel (SOL) and solid state (SS) techniques using rigorous quantitative measurements to separately compare material diffusivities and galvanostatic performance. Here, the SOL samples had shorter length-scales and lower diffusion coefficients. Galvanostatic cell measurements, however, revealed that the shorter length-scales more than compensated for the lower diffusivities with better overall high-rate capacity retention. This example shows how cell level metrics often differ from underlying diffusivities. We argue that materials development needs renewed focus on property measurements like diffusivity where best-practices are important to derive meaningful insights towards structure-property relationships.

## Introduction

The fast-moving field of battery research is evolving to bring the required versatility for energy storage devices through the discovery and enhancement of novel active materials.[1–11] Structure-property relationships are foundational and are often combined along with processing and performance as the corners of the "materials science tetrahedron." In the context of batteries, performance is often limited by solid state ionic diffusion which is in turn constrained by the material property of diffusivity as well as the material length scale. Diffusivities are often calculated from cell performance data (e.g. current/voltage transients) using either the nominal particle radius or the mass specific surface area, both of which are conveyed in our broader term "material length scale."[12,13] The authors were motivated to write this perspective to help the community establish strong structure-property claims for diffusivity that follow best practices.

Field trends were estimated by analyzing 303 open-access publications on the topic of battery active materials development from the past five years. A large language model was trained to interpret the reported experimental procedures and the scope of claims made in each manuscript.

For example, galvanostatic cycling and current/voltage transient measurements of transport were separately identified. Methods of assessing material length scale were identified, including physisorption (Brunauer-Emmett-Teller), microscopy (electron, optical), and X-ray methods (Scherrer, Porod).[3,14–18] The specific sequence of sample preparations was parsed with particular attention to the timing of grinding, e.g. occurring before length scale measurement vs after. Often the sequence of these events was ambiguous despite the significant impact on the apparent diffusivities, *vide infra*. Figure 1 presents a summary of this analysis. The analyzed publication set had ~96% new-reports with roughly equal fractions of experimental-only and combinations of experimental with computation (Fig 1a). More than 90% of the publications proposed structure-property relationships, highlighting their perceived value (Fig 1b). Of the publications noting structure-property relationships, there was roughly equal proportions of those reporting cell level metrics alone (e.g. galvanostatic rate capability) versus those that included diffusivity values. It is important to keep in mind that property measurements, such as diffusivity, are of course required to establish property relationships.

Next attention in the publication analysis was turned to how diffusivity values were determined. As a motivating example, $Ti_2Nb_{10}O_{29}$ diffusivities were determined in 6 of the analyzed publications with values ranging over 5 orders of magnitude from $\sim5\times10^{-19}$ to $\sim1\times10^{-13}$ $m^2s^{-1}$. How much of this variation is due to structural differences (defects) versus analytical method differences? The sequence of analytical events, e.g. grind-measure vs measure-grind were parsed where the majority of publications did not clearly state this sequence of events ("ambiguous") as shown in Fig 1d. As will be demonstrated with example data later, changing this sequence substantially changes the apparent diffusivity values. The diversity of methods for length scale assessment were quantified with similar fractions for physisorption, microscopy, and X-ray methods (Fig 1f). Many publications included multiple length scale methods, but only ~57% of diffusivity reports clearly stated the length scale used in diffusivity calculations (Fig 1e). The LLM was trained to categorize using explicit statements without presumptive inference which partially explains the large fraction of "ambiguous" categorizations. This brief analysis of recent publications demonstrates that transport-related structure-property claims deserve refocused attention.

Battery measurements for the purpose of material property assessment are ideally free from the contributions of the electrode and cell design (Fig 1g). That is, simplicity is preferred with e.g. half-cell measurements and, furthermore, competing rate constraints should be minimized from the counter electrode, electrolyte, and electrode design (additives, porosity).[19–21] For example, the use of oversized lithium chips (0.6 mm thick) avoids issues with depleting the lithium inventory during cycling. Likewise, the use a few $mg/cm^2$ of active material minimizes polarization from electron and electrolyte transport.[22] For comparisons with work elsewhere, additional factors like electrolyte concentration, carbon additives, and binders can be constrained to conventional conditions (e.g. 1.0 M electrolyte, 80:10:10 ratio of active material:carbon:binder) for broader comparability.[23] Designing battery measurements in these ways makes the electrochemical data largely reflect the diffusivity in the active material of study. Galvanostatic C-rate capacity retention is a common comparison metric for battery studies, however, this convolves effects from material property changes with particle size changes and other cell-level effects from electrolyte, additives, binders, cell configuration, loading thickness, as well as environmental factors (Fig 1g).[24] Electrodes made with the same material can have markedly different C-rate capacity retention stemming from differences from any of these variables where the material itself did not necessarily change its diffusion in any way.[25,26]

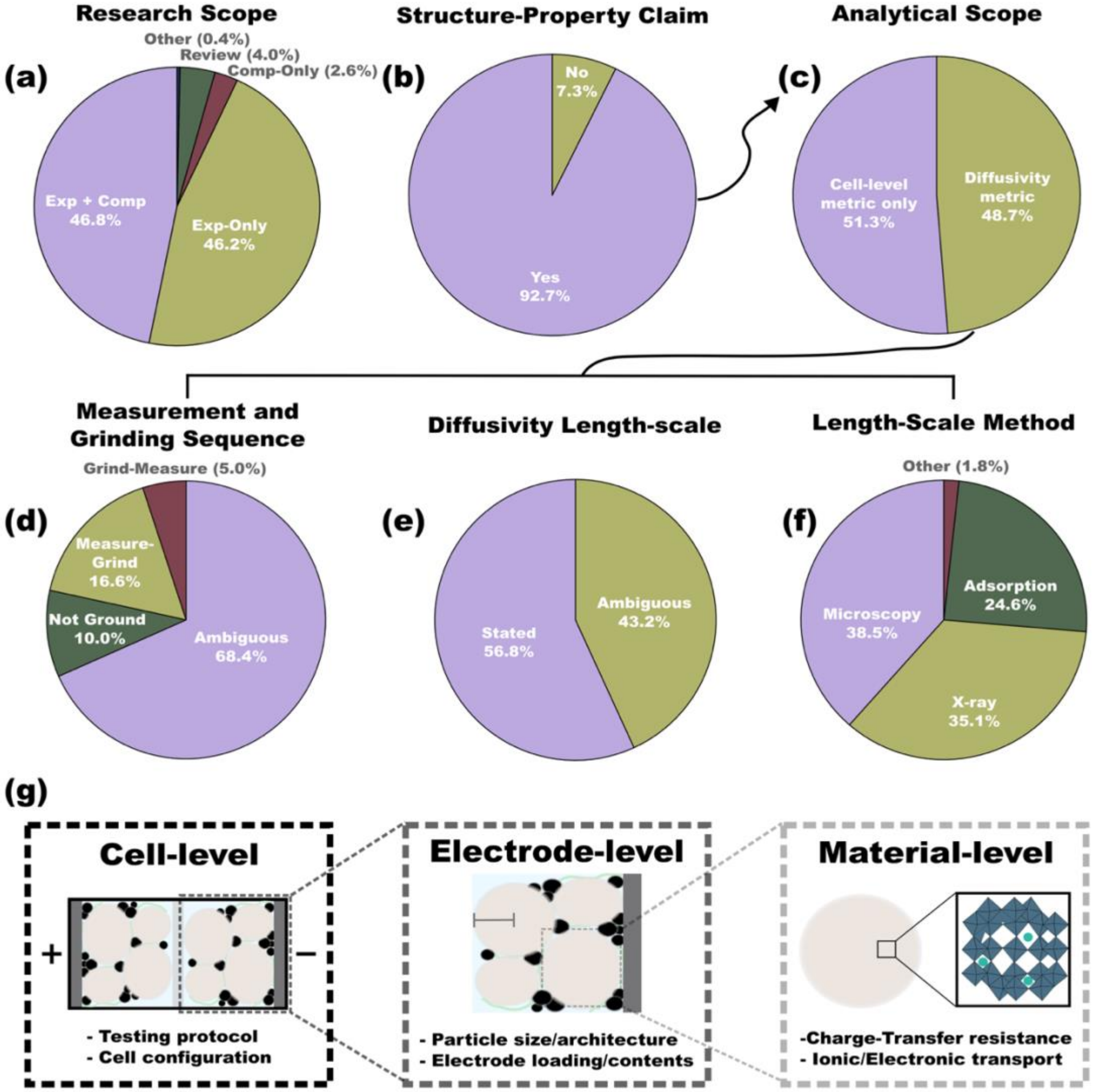

Figure 1. A total of 303 open-access publications about battery materials from the past five years were analyzed. The (a) research scope and (b) inclusion of structure-property claims were parsed. From the subset of publications with structure-property claims, the (c) analytic scope, the (d) sequence of grinding and length-scale measurements, the (e) clarity of length scale used in diffusivity calculations, and the (f) diversity of length-scale measurement techniques were identified. A (g) schematic showing the gap between cell-level metrics with many convolved contributors and material-level metrics.

A variety of common voltage/current transient techniques are used for determining diffusion coefficients. For example, intermittent current interruption (ICI), galvanostatic intermittent titration technique (GITT), potentiostatic intermittent titration technique (PITT), and electrochemical impedance spectroscopy (EIS) are common.[27–33] The charge storage process always involves multiple simultaneous process steps (Fig 2a) where the time dependence of the response is fundamental to identifying diffusion-limited responses ($t^{0.5}$). For example, with ICI the V(t) profile during current interruptions can be separated into several different overpotential contributions.[28,34] For brevity sake, those are presented here as a diffusion overpotential ($\Delta V_D$) and a non-diffusion overpotential ($\Delta V_{ND}$) in Fig 2b, as described elsewhere in detail.[24,28,34–36] Regardless of which of these techniques is used to determine diffusivity, these all require the material length-scale (Fig 2c) to estimate the electrochemically active surface area.[12,13,24,37–39] This length scale is conventionally implemented via the mass specific surface area (A) or the nominal particle radius (r). Measuring the true electrochemically active surface area for batteries, however, remains a grand challenge since the active material is partially coated with binder and carbon additives.[31,40,41] Such mixed and partial coatings inhibit accurate assessment of uncoated active material via typical physisorption or X-ray methods. Active materials are often hierarchical with large secondary particles composed of aggregates of smaller primary particles. The secondary particle surface area was suggested to be used as the electrochemically active surface, however this presumes that the secondary particle surfaces are uncoated and the primary particle surfaces are completely covered by additives.[40] Due to analytical complexity, however, this suggestion has remained without validation. On the other hand, the conventional use of particle radius or total mass specific surface area presumes that all active material surface area is electrochemically active which is an overestimation. While both approaches are imperfect and result in an "apparent diffusivity," we argue that the latter is preferable by having fewer assumptions. Regardless of the method utilized, the combination of length-scale information with electrochemical measurements enables calculation of the key material property of diffusivity.

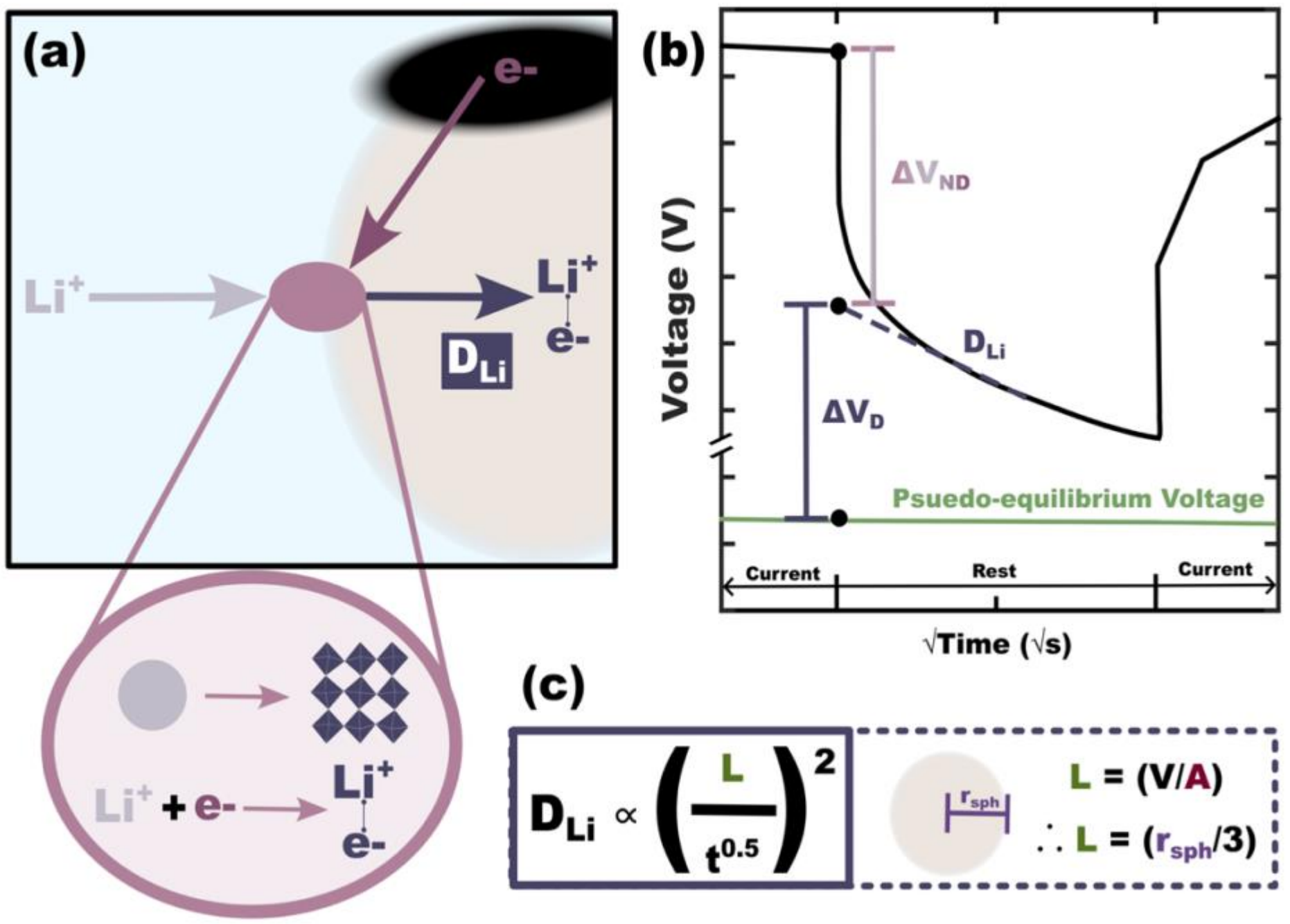

Figure 2. Overview (a) diagram of processes occurring in a battery electrode consisting of a single spherical particle of active material, including ionic electrolyte transport, electron transport, charge transfer, and solid-state ion intercalation. Intermittent current interruptions yield (b) voltage relaxation profiles. Comparison to pseudoequilibrium potentials enables the deconvolution of diffusion ($\Delta V_D$) and non-diffusion ($V_{ND}$) overpotentials. Translating these electrode-level measurements to material-level (c) diffusivity calculation requires the material length-scale (L) which is generally implemented as a nominal particle radius (r/3) or the V/A ratio based on the mass specific surface area.

Not all length scale information is equally valid, especially when different grinding procedures are used for battery production versus length scale measurements. Figure 3 presents two distinct experimental procedures with different sequences of steps. The left side depicts the measure-grind strategy that was found in 16.6% of the 303 publications analyzed above. Here the active material is at some point ground with additives after the assessment of the active material length scale. Clearly this changes the active material length scale, adding further uncertainty to any derived diffusivities as will be demonstrated later. Unfortunately, ensemble techniques for mass specific surface area measurements of such composites do not yield the electrochemically active surface area since the additive surface area is convolved. The right side of Figure 3 presents a grind-measure strategy that was found in 5% of the 303 analyzed publications above. Here a single grinding step is used for the active material-alone before length scale measurement and is furthermore without subsequent grinding during battery preparation. For example, subsequent mixing with carbon and binder solution can occur via stirring so that the active material length-scale is not altered further. Notably 68.4% of the 303 analyzed publications had experimental procedures that were not sufficiently detailed to identify the sequence of grinding and measurements. For accuracy's sake, we advocate that the community exclusively use the grind-measure strategy for diffusivity data and document it with sufficient experimental details.

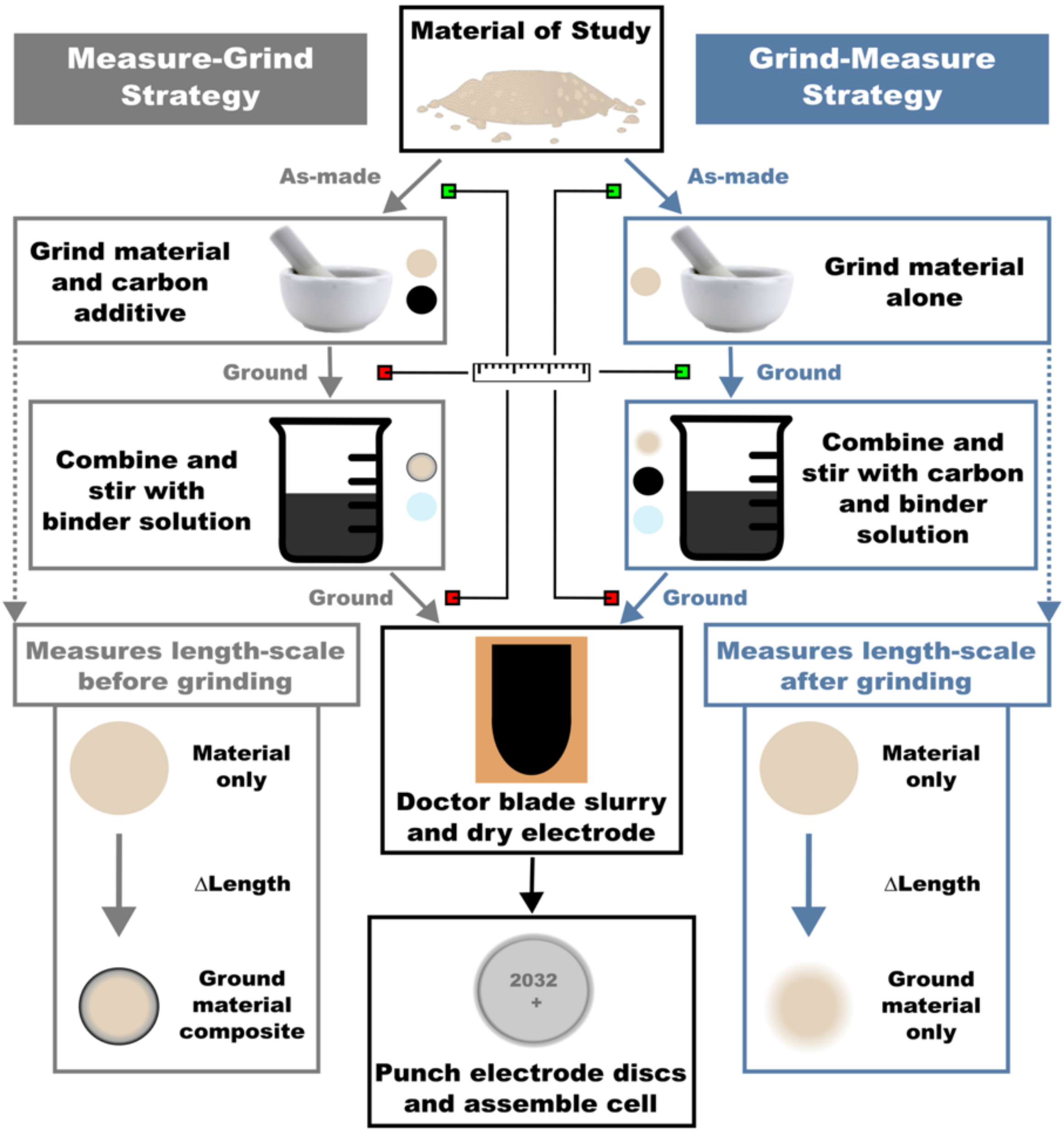

Figure 3. Schematic comparing the measure-grind and grind-measure strategies. Grinding is generally necessary to reduce aggregation and improve kinetics, however, grinding with additives hampers the subsequent measurement of the material length scale since the additives also contribute to mass-specific surface area. On the other hand, the grind-measure strategy includes grinding the pure active material so that its final length scale is measurable alone. Importantly, the subsequent electrode production with the grind-measure strategy should not include further grinding so that the material length scale remains accurately known.

In this perspective we advocate for best practices for battery material diffusivities with consideration to material length-scale assessment. Experimentalists should be careful that the active material length scale used in batteries is identical to that measured separately and clearly describe the methods used to do so. Length-scale measurement techniques such as SEM, BET, and SAXS Porod analysis are compared and discussed in detail, including examples of how errors significantly alter the reported apparent diffusivities for the same sample. With diffusivity values changing substantially with state of charge, we also advocate for figure-of-merit based comparisons such as capacity-weighted diffusivity that combine all diffusivity data into a single representative value.[42] Example cases are elaborated, clarifying the difference between material-level and cell-level metrics. The example materials include $TiNb_2O_7$ (TNO1) and $Ti_2Nb_{10}O_{29}$ (TNO2) which were both synthesized by solid state (SS) and sol-gel (SOL) methods. As a final example, cell-level and material-level metrics are compared by their apparent rankings for the four samples, demonstrating how convolved length-scale changes effect properties and hinder apt structural comparisons.

## Experimental

*Large Language Model Analysis of Publications*

A corpus of 303 battery research articles was compiled from the Scopus database (Elsevier), consisting of open-access publications from the past five years that were identified as focusing on either cathode or anode materials development. Experimental procedures were extracted using OpenAI Codex through a structured, prompt-based information extraction workflow executed locally. The workflow consisted of a predefined set of analysis criteria, decision rules, and reporting conventions that were iteratively refined to improve consistency and reproducibility across the dataset. This workflow was applied uniformly to all analyzed publications. The reliability of the resulting analysis was validated manually using a subset of the articles.

*Chemicals*

Titanium tetraisopropoxide (TTiP, Sigma-Aldrich, 99.99%), niobium ethoxide (NbOEth, Sigma Aldrich, 99.5%), nitric acid (Sigma-Aldrich, 70%), acetic acid (Sigma-Aldrich, 99.7%), ethanol (Decon Laboratories, 100%), Super P Conductive Carbon Black ("carbon black" MSE), N-methyl-2-pyrrolidone (NMP, Sigma-Aldrich, 99.5%), poly(vinylidene fluoride) (PVDF, MTI, 99.5%), acetone (Fisher,99.5%), lithium hexafluorophosphate in 1:1 ethylene carbonate/dimethyl carbonate (Sigma-Aldrich), lithium metal (MSE, 99.9%), $TiO_2$ (anatase, Alfa Aesar, 99.9 %) and $Nb_2O_5$ (thermo scientific, 99.9985%) were used as received.

*Solid State Synthesis*

To synthesize $TiNb_2O_7$ and $Ti_2Nb_{10}O_{29}$ using the solid state reaction approach, stoichiometric quantities of $Nb_2O_5$ and $TiO_2$ were intimately mixed in an agate mortar and pressed into pellets using 2 tons of pressure in a hydraulic press. The pellets were placed into an alumina crucible, covered with a lid and heated in a high temperature furnace. For $TiNb_2O_7$, the sample was heated at 1350 °C for 3 days with intermittent grindings. In the case of $Ti_2Nb_{10}O_{29}$, the pellet was heated at 1200 °C for 3 days with intermittent grindings. Upon completion of the reaction, the sample was characterized by powder X- ray diffraction.

*Sol-Gel Synthesis*

The sol-gel-based $TiNb_2O_7$ (TNO1-SOL) was produced by spray drying a 1:2 molar ratio of Ti precursor (TTiP) and Nb precursor (NbOEth) as described elsewhere in detail.[26] Following the spray drying, TNO1-SOL was heat treated at 800°C for 2 hrs with a 5°C/min ramp rate and natural cooling. The sol–gel sample of $Ti_2Nb_{10}O_{29}$ (TNO2-SOL) was synthesized using titanium tetraisopropoxide and niobium ethoxide as precursors. To synthesize it, 71.05 mg of TTiP, (0.250 mmol) and 397.73 mg of $Nb(OEt)_5$, (1.2499 mmol) were dissolved in 3.0 mL of ethanol. The mixture was stirred for 5 minutes to obtain a homogeneous solution, then 0.5 mL of acetic acid was added and stirred. Deionized water (0.5 mL) was added dropwise to initiate gel formation. The gel was heated at 100 °C overnight to remove ethanol and water. The dried powder was calcined at 1200 °C for 6 h with a heating ramp of 5 °C $min^{-1}$, then allowed to cool naturally to room temperature.

*Scanning Electron Microscopy*

Scanning Electron Microscope (SEM) images were acquired using a Zeiss Gemini500 with an accelerating voltage of 5 keV and an in-lens detector for secondary electrons. The images were acquired with a working distance of 2.8 mm at magnification settings of 55KX or 180KX. ImageJ software was utilized for quantitative image measurements. The effective radius was measured from the span of solid between pores (2*r). The mean values are reported with standard-error-of-the-mean based on the number of analyzed radii.

*Gas Physisorption*

Brunauer-Emmett-Teller (BET) method measurements were performed using a Micromeritics 3Flex instrument. The sample was first degassed under $N_2$ flow at 300 °C for 24 hrs. Nitrogen (77K) isotherms were measured over a relative pressure ($P/P_o$) range of 0.004-0.500 while the BET method assessment was applied on a range if 0.060-0.300 on the standard cross-sectional area of 0.162 $nm^2$. The resulting isotherms were analyzed using 3Flex software as displayed in Figure S1. The best-fit value is reported with the error of the fit.

*X-ray Scattering*

Small-angle X-ray scattering (SAXS) and Wide-angle X-ray scattering (WAXS) measurements were used for Porod surface area analysis and phase identification, respectively. Measurements were performed on a SAXSLab Ganesha at the South Carolina SAXS Collaborative (SCSC). A Xenocs GeniX 3D microfocus source with a copper target to generate a monochromatic X-ray beam with a wavelength of 0.154 nm and an intensity of $5.54\times10^7$ counts per second was used. The scattering angle of the instrument was calibrated prior to WAXS measurements using the National Institute of Standards and Technology (NIST) reference material 640d silicon powder with a peak position of $2\theta = 28.44°$, where $2\theta$ is the total scattering angle. A Pilatus detector captured the scattered signal and was also used to measure the direct/transmitted beam intensities. Samples were mounted using Kapton tape and all data had the Kapton background signal subtracted using SAXSGUI software. The scattering data is presented where the magnitude of the momentum transfer vector q corresponds to $q=4\pi\sin(\theta)/\lambda$. Standardless absolute scattering intensities were reported with the sample thickness determined using the "apparent thickness" approach developed by Spalla et al.[43,44] The X-ray absorption coefficients of 462 $cm^{-1}$ and 463 $cm^{-1}$ (NIST calculator) were used with the measured transmissivities to calculate the apparent sample thicknesses of TNO1 and TNO2 samples, respectively.[45] Importantly, the X-ray collimation was kept constant during the transmittance measurement so that the illuminated region was the same as that used during the SAXS measurement (powders have heterogeneous thickness). Furthermore, the counts/second for each pixel of the Pilatus detector were validated to be within the linear regime for quantitative accuracy. Plots were used to calculate the specific surface area starting from the surface to volume ratio defined by:

$$\Sigma = \frac{(I_{abs}q^4)}{2\pi(\Delta SLD)^2}$$

Where ΔSLD was equal to the scattering length density of the material. The SLD value of $3.38\times10^{-5}$ $A^{-2}$ for $TiNb_2O_7$ (TNO1) and $3.46\times10^{-5}$ $A^{-2}$ for $Ti_2Nb_{10}O_{29}$ (TNO2) was calculated using a NIST resource for the utilized X-ray wavelength used.[46] The measurable Porod region was indicated as the constant $I_{abs}q^4$ portion of the scattering data where the high-q limit was constrained to values with less than 1% intensity error. The sample specific surface area (S) was then calculated using:

$$S = \frac{\Sigma}{\rho}$$

where **ρ** is the sample bulk density (4.295 g/cm$^3$ and 4.415 g/cm$^3$ respectively for TNO1 and TNO2). The specific surface area for each sample was measured for the as-made powders and after grinding in a mortar and pestle. The resulting Porod scattering profiles with q-ranges utilized are presented in Figure S2. Porod surface areas are derived from the mean S value with the standard-error-of-the-mean based on the number of analyzed q-points.

*Electrochemistry*

Electrodes were prepared from slurries for lithium half-cell measurements. Each active material was individually ground with a mortar and pestle where an aliquot was collected for surface area analysis. The ground material was then combined with carbon black in a dram vial at an 8:1 ratio of active material to carbon. Following that, a 25 mg/ml solution of PVDF in NMP that was added to create an overall ratio of 8:1:1 of oxide, carbon black, and PVDF. The NMP-based mixture was stirred for an hour to ensure thorough mixing. Importantly, this slurry was not ground further since that would alter the active material length scale. Copper foil current collectors were cleaned by wiping using Kimwipes moistened with acetone. The slurry was doctor bladed onto the copper foils using a blade height of 15 μm and were immediately dried on a pre-heated hot plate set to 85 °C for 20 min. The mass loading for all electrodes were between 1-2 mg/cm$^2$.The films were further dried in a vacuum oven set to 100 °C for 20 hrs. After drying, the electrodes were punched to 12 mm discs, weighed in triplicate, and brought into an argon glovebox for assembly into coin cells. Coin cells were assembled using a 2032-type cases/lids, 1.0 M $LiPF_6$ in ethylene carbonate/dimethyl carbonate electrolyte, Whatman glass fiber separators, 16 mm lithium metal chips, two 0.5 mm stainless steel spacers, and 1 stainless steel wave spring. A Biologic BCS-810 was used for galvanostatic cycling and intermittent current interruption (ICI).[27] The cells were tested in a room at 21°C and each started with constant current cycling at 0.05C for five cycles before completing ICI. The ICI technique was used to calculate diffusivities and overpotentials with a current density corresponding to 0.1C.[5] A period of current was applied for 300 s, followed by a zero current interruption for 10 s. These interruptions were repeated throughout the voltage range of 1.0-3.0V vs $Li/Li^+$. Detailed overpotential measurements utilized the Bio-Logic modulobat technique to capture voltage transients with minimal readout delay. The ICI method calculates each diffusivity value (D) as:

$$\boldsymbol{D} = \frac{4}{\pi}\left(\frac{V}{A} * \frac{\frac{\Delta E_{OC}}{\Delta t_1}}{\frac{dE}{dt^{0.5}}}\right)^2$$

where V is the molar volume of the electrode material, A is the electrochemically active surface area, $E_{OC}$ is the open circuit potential, $\Delta t_I$ is the period of constant current applied between OCP measurements, E is the potential of the electrode, and t is the step time. A root-time fitting of 1-5 s was selected for $D_{Li}$ calculations and was validated for consistency with linearity.[27] During the transient current pause, the difference between the t=0 extrapolated diffusion potential and the potential just prior to the current pause were termed the non-diffusive overpotential ($\Delta V_{ND}$).[47] Diffusion overpotential ($\Delta V_D$) was calculated by comparison to a pseudoequilibrium 0.02C GCD measurement where the minor hysteresis was accounted for by averaging the lithiation and delithiation voltages at each $Li_x$ extent.[34,48] C-rates were defined based on one electron per transition metal, corresponding to a 1C current density of 232.6 mA/g for $TiNb_2O_7$ (TNO1) and 216.0 mA/g for $Ti_2Nb_{10}O_{29}$ (TNO2). Galvanostatic cycling was conducted between 1.0-3.0 V vs $Li/Li^+$ with a 1 hr voltage hold period after delithiation. Each condition was measured in triplicate to present mean battery values along with the error-of-the-mean. Capacity-weighted metrics were calculated for each sample individually before determining the sample-to-sample mean and error-of-the-mean (n=3 for triplicates). The specific surface areas used in diffusivity calculations were derived from SAXS Porod analysis using the Grind-Measure approach.

**Results and Discussion**

First we show how the active material length-scale compares when measured via electron microscopy, physisorption, and X-ray techniques. Scanning electron microscopy (SEM) is widely available and directly yields insights into the active material morphology and size. As a surface imaging technique, SEM data corresponds to the exterior porosity of the sample where sample fracturing is sometimes used to reveal interior porosity. Regardless, microscopy data are localized (non-ensemble) and require rigorous sampling to avoid user bias. Furthermore calibration is needed to avoid systematic error which varies with working distance and magnification.[49] With typical samples containing heterogeneity, numerous repeated measurements are important to assess mean values and the standard error-of-the-mean (SE). The selection of magnification can also introduce user bias. For example, SEM data for TNO1-SOL was analyzed with two magnifications (Fig 4a,c). The lower magnification micrograph emphasized the secondary particle size (Fig 4a,b) whereas the higher magnification micrograph included the primary particle sizes (Fig 4c,d). The resulting histograms (Fig 4b,d) vary significantly. The size-heterogeneity is apparent in the large standard deviations that corresponded to 39-63% of the mean where the inclusion of 150 measurements significantly reduced the SE to 3.2%. Hypothetical under-sampling of the same Figure 4c micrograph with 5 or 25 measurements here would yield mean values respectively with 18% or 8% SE that would lead to substantial diffusivity uncertainty when propagated (36% or 16% error). Reports also vary on the usage of the longest vs shortest particle axes when measuring anisotropic shapes where randomized sampling orientation is suggested as an unbiased approach. Both physisorption and X-ray Porod analysis can provide ensemble measurements of the mass specific surface area. Gas physisorption isotherms are the most popular route to determine the specific surface area where this data is usually analyzed with the Brunauer-Emmett-Teller (BET) model that presumes non-specific and weak interactions. Notably the quadrupole moment of $N_2$ is not ideal for polar oxide surfaces and leads to increased error.[50] In contrast, Ar is a more ideal adsorbent where the resulting BET analysis can have a high degree of accuracy with less than 1% error.[17,50,51] There are additional possible errors from model selection and analysis methods where a round robin study including more than 50 labs analyzing the same 18 isotherms found 7-31% difference in the interpreted specific surface areas.[52] Also, the typical instrument requirement for ~1 $m^2$ of surface area corresponds to a gram of sample when the surface area is often ~1 $m^2$/g. This large sample quantity poses a challenge for small-scale and many-sample studies. Lastly, Small-Angle X-ray Scattering (SAXS) is another ensemble technique for measurement of mass specific surface areas. Specifically, Porod analysis of absolute intensity scattering data enables the direct calculation of mass specific surface area. The measured Porod surface area includes all surfaces (oxide-vacuum interfaces) includng both primary and secondary particles. The most challenging aspect of this SAXS method is accurate measurement of the material thickness which was elegantly addressed by Spalla through the use of transmittance measurements with knowledge of material's X-ray attenuation coefficient to determine the apparent thickness.[44,53] Though SAXS instruments are less widely available, this approach only requires e.g. ~10 mg of material for transition metal oxides and can be completed within a couple hours with accuracy comparable to Ar BET.[17,42,54]

The surface areas and diffusivities resulting from each of these techniques are next compared using sample TNO1-SOL. Here, the $N_2$ BET and Porod SAXS analysis resulted in specific surface area values that were statistically indistinguishable whereas the SEM primary particle method was 17% higher, possibly due in part to instrument calibration (Fig 4e). With ideal conditions, all three of these techniques should provide similarly accurate values, however, the ensemble methods are naturally the most reliable due to the volume of sample analyzed. Notably, the secondary particle size measured by SEM was an order of magnitude lower than the other measures. Each of these length-scale measurements were used to calculate apparent diffusivities for TNO1-SOL.[42] Here the same electrochemical data was analyzed (same batteries) repeatedly using these different measured material length-scales (Fig 4f). As expected, the similar surface areas from BET and SAXS Porod methods on as-made materials led to similar apparent diffusivities. Correspondingly, the SEM primary particle measurements with higher surface area led to lower apparent $D_{av}$, whilst the SEM secondary particle measurements led to significantly higher apparent $D_{av}$. Importantly, however, none of these measurements on as-made materials accounted for the length-scale change that occurs with the grinding used for battery production. SAXS Porod analysis was repeated after grinding following the grind-measure strategy which led to an apparent diffusivity of $2.09\times10^{-19}$ $m^2s^{-1}$. This diffusivity value was calculated from triplicate cells with a standard deviation corresponding to 19% of the mean. High-current density cycling can also cause cracking that changes the electrochemically active surface area during cycling which does not have an apparent resolution in terms of accurate diffusivity measurements.[55] In this light, diffusivity measurements should be carried out on cells without inducing such length scale changes. Back to the main point, length scale metrics (both primary and secondary particle sizes) obviously change with grinding where materials studies should rationally follow the grind-measure strategy.

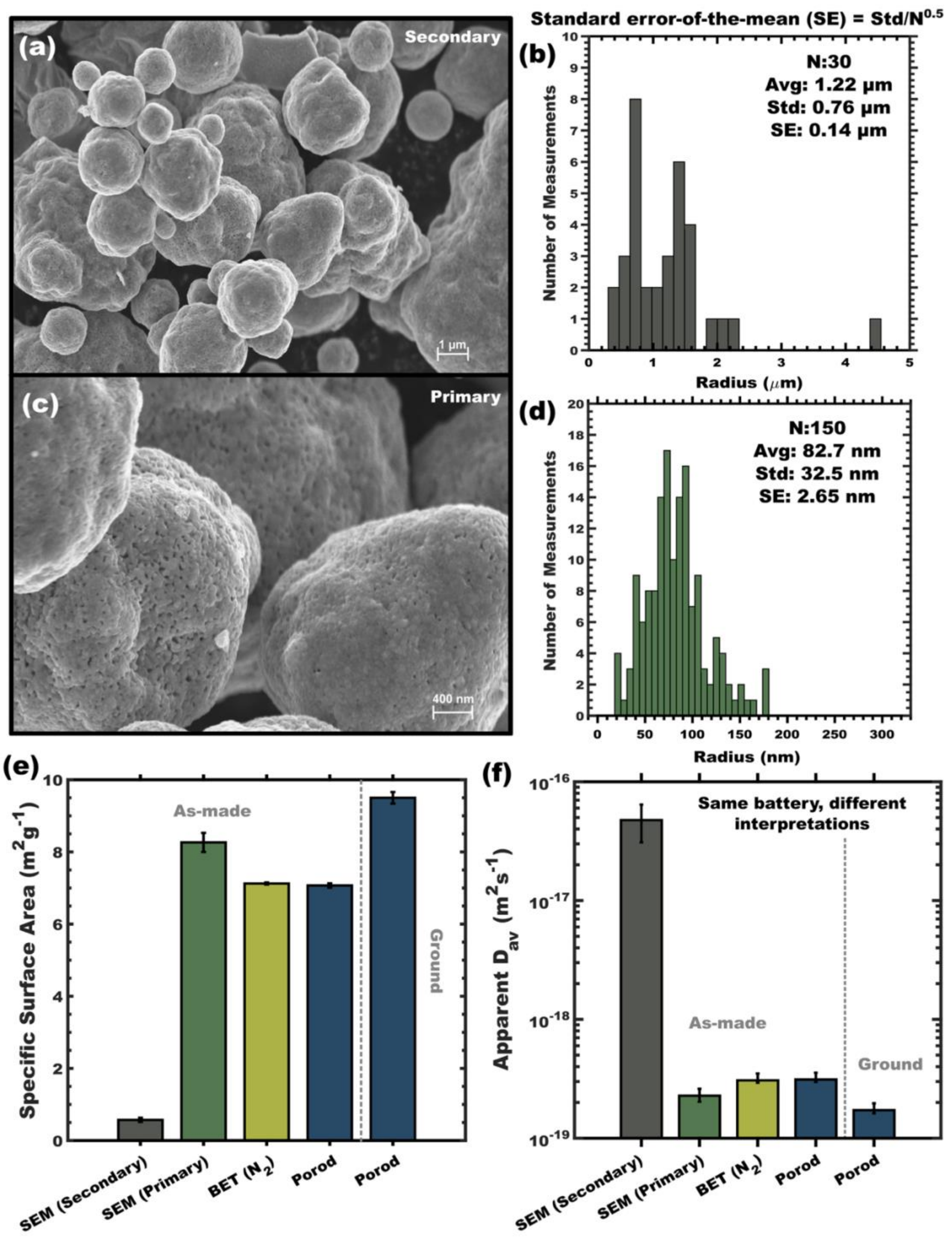

Figure 4. SEM micrographs (a,c) of as-made TNO1-SOL at different magnifications along with the size histograms (b,d) reflecting the secondary and primary particle sizes. Length scale measurements using SEM, BET, and Porod were compared using mass specific surface area (e), including Porod analysis after grinding. A single electrochemical dataset (3 identical batteries) were analyzed using these different length scale measurements to (f) compare the different apparent diffusivities. Error bars correspond to standard deviation for each method.

Four different TNO samples were next compared using the best-practices advocated for in this perspective. To reiterate the earlier point about grinding effects, the surface areas are shown to all change significantly from grinding with about an order of magnitude change in one case (Fig 5a). Notably TNO1-SOL had the highest specific surface area 9.50 $m^2/g$ that was produced via a low temperature spray-drying technique. The ICI overpotentials were compared where each sample had about an order of magnitude higher diffusion overpotential ($\Delta V_D$) than all other overpotentials combined ($\Delta V_{ND}$) (average data in Figure 5b, full dataset in Figure S4). A Li-Li symmetric cell was examined with a current density corresponding to the 0.1C ICI measurements here where the counter electrode contributed an insignificant ~2-3 mV of overpotential (Fig S5). The D(x) profiles were calculated for each sample using the grind-measure data (Fig 5c). Peaks in D(x) profiles have been connected with lithium ordering where the flatter profile for TNO1-SOL is consistent with its expected defect-rich character from low temperature synthesis.[34,56]Whereas comparison of arbitrary single point diffusion values vary the sample rankings due to cross-over points, capacity-weighted average diffusivities ($D_{av}$) are figures-of-merit that enable sample comparison regardless of D(x) complexity (Fig 5d).[42] Here the solid state synthesized samples (TNO1-SS and TNO2-SS) both have the highest capacity-weighted average diffusivity of $4.97\times10^{-16}$ $m^2s^{-1}$. These high diffusivities correlated with higher crystallinity provided by the solid state route. For example, the disorder of TNO1-SOL is apparent in its broadened diffraction peaks (Fig S3) as examined elsewhere in detail.[26] On the other hand, TNO2-SOL was crystallized with a similar temperature as the solid state samples and had similar capacity-weighted average diffusivity of $2.55\times10^{-16}$ $m^2s^{-1}$. These four samples can be ranked in terms of the material property of lithium diffusivity for (TNO2-SS=TNO1-SS>TNO2-SOL>TNO1-SOL). As noted in the analysis of 303 publications, 92.7% of these battery articles developing materials made a structure-property transport claims with about half using diffusivity measurements and about half using cell-level metrics such as galvanostatic capacity retention. Having established the diffusivity ranking above, galvanostatic capacity retention is next compared to see if that is a suitable proxy for diffusion measurements.

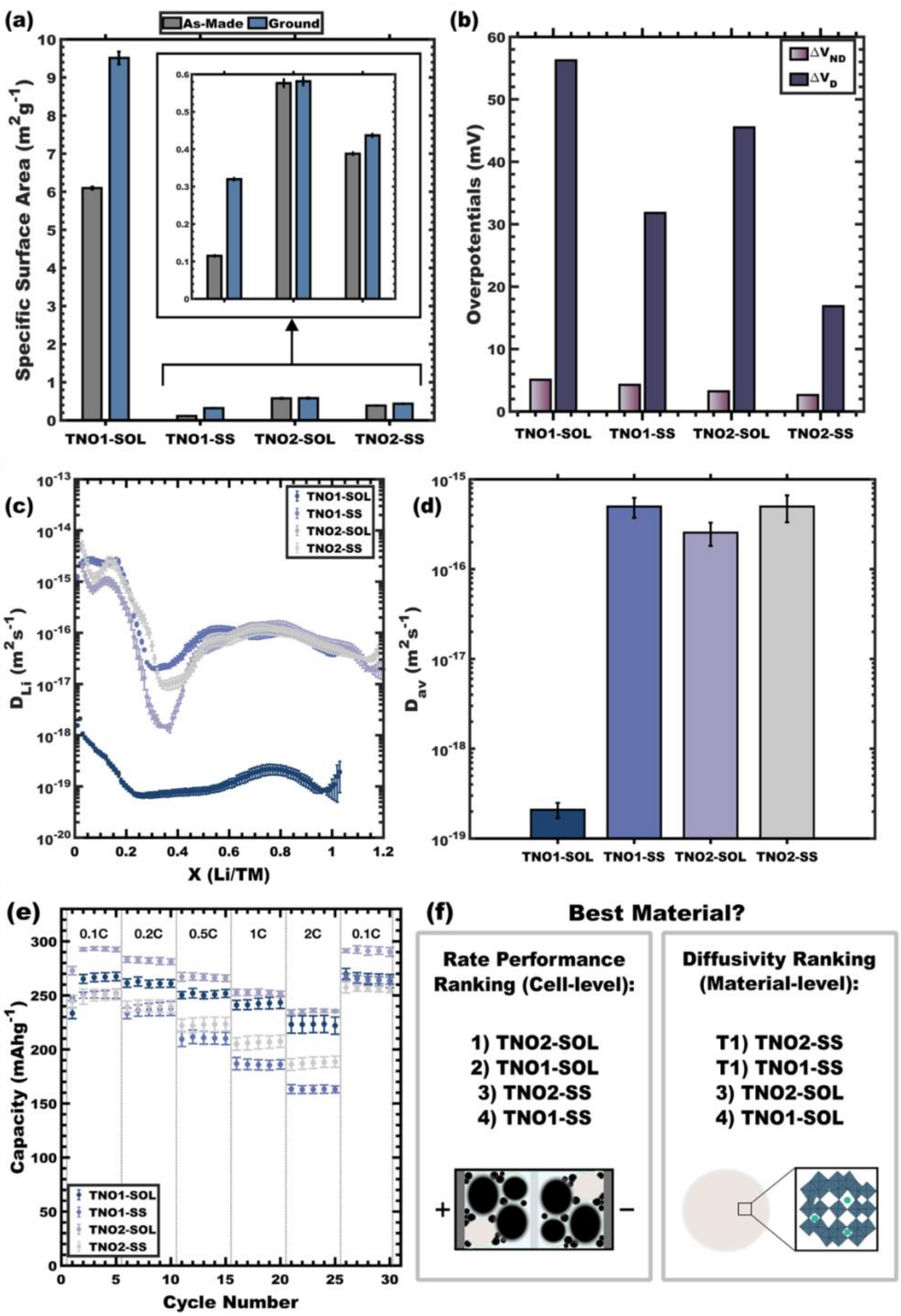

Figure 5. A (a) comparison of each sample's surface area determined by Porod analysis on as-made and grind-measure samples. Capacity-weighted average overpotentials are shown for the (b) non-diffusion and diffusion overpotentials for all four samples. The (c) full apparent diffusivity profiles are shown as well as their (d) capacity-weighted average diffusivities. Galvanostatic capacities (e) at varying C-rates for each sample. A (f) comparison of rankings based on either cell-level capacity retention or material-level diffusivity. All error bars correspond to the standard error-of-the-mean from battery triplicates.

The C-rate capacity retention results for these samples is shown in Figure 5e. As usual, the capacities decrease with increasing current density as the overpotentials progressively increase and limit the amount of charge passed within the voltage window. At 1C and 2C, the capacities (Figure 5e) show a clear trend with TNO2-SOL>TNO1-SOL>TNO2-SS>TNO1-SS. In this particular example, the ranking from galvanostatic capacity retention is nearly the opposite of the material diffusivity ranking (TNO2-SS=TNO1-SS>TNO2-SOL>TNO1-SOL, Fig 5f). Synthetic changes usually alter both the diffusivity and the as-made active material length scale which clearly can obscure whether or not the material itself improved (Fig 1g). Thus, while galvanostatic rate capability does provide a practical demonstration of cell performance, it is not an appropriate method to compare material properties (Fig 5f). It is also remarkable that TNO1-SOL maintains such high capacities at high C-rates despite its relatively poor diffusivity. This is perhaps an underappreciated opportunity to better understand how to scale down materials while preserving high diffusivity. Notably the sequence of grind-measure vs measure-grind does not alter the galvanostatic performance since these only vary the timing of the length scale measurement when limited to grinding the active material alone. There may be effects of grinding with additives on galvanostatic performance, however that is outside the scope of this study. Overall, a variety of methodologies were discussed to suggest best-practices for ranking battery active materials based on accurate assessment of their diffusivities.

## Conclusion

The demand for improved batteries ultimately connects to a need for improved active materials. Performance metrics for materials are critical for further development via comparative-ranking. A review of 303 recent battery material publications revealed (1) an emphasis on structure-property transport claims about battery materials with only about half of those studies measuring diffusivity and (2) that of the studies reporting diffusivity, 15% clearly articulated experimental procedures such as grind-measure that avoid the systematic error of measure-grind approaches. Best practices were suggested for diverse methods of measuring active material length scale where advocation was directed at improved experimental clarity for grinding steps. Common errors and pitfalls of popular length-scale measurement techniques were noted. Example quantitative comparisons were made using two active material compositions, each prepared with two synthesis methods. Generally, the solid state samples had longer length-scales and higher diffusion coefficients. Cell-level capacity retention measurements, however, revealed that the smaller particle sizes of the sol-gel samples more than compensated for their lower diffusivities which enabled better high-rate capacity retention. This data also demonstrates that capacity retention data is insufficient to rank diffusivity changes since the active material length scale has substantial influence on diffusion polarization. We hope that future battery material studies will ensure that (a) structure-property transport claims include diffusivity measurements, (b) a strategy such as grind-measure is used to match measured particle surface areas to the particles used in batteries, and (c) experimental procedures should clearly present the timeline of particle grinding and measurements.

## Data availability

Data for this perspective, including galvanostatic data and ICI data for all four samples are available at Open Science Framework at osf.io/5a4mw.

**Conflicts of Interest**
The authors declare no conflict of interest.

**Acknowledgements**
The authors gratefully acknowledge the support from the U.S. Department of Energy (DOE), Office of Basic Energy Sciences, under Award #DE-SC0023377. This work made use of the South Carolina SAXS Collaborative.